\documentclass[review]{elsarticle}
\usepackage{epsfig}
\usepackage{a4wide}
\usepackage{graphicx,amssymb}
\usepackage{amsmath
}        

\def\lb#1{\if 1#1 \ln\beta \else \ln^#1\beta \fi}
\def\lt#1{\if 1#1 \ln 2 \else \ln^#1 2 \fi}

\newcommand{\be}{\begin{equation}}
\newcommand{\ee}{\end{equation}}
\newcommand{\ba}{\begin{eqnarray}}
\newcommand{\ea}{\end{eqnarray}}

\begin{document}

\vspace{\baselineskip}

\title{
The $\langle VVA \rangle $ correlator at three loops 
in perturbative QCD
}

\author[kgc]{ Jorge~Mondejar}
\author[km]{Kirill~Melnikov} 

\address[kgc]{Institut f\"ur Theoretische Teilchenphysik, 
University of Karlsruhe, Karlsruhe, Germany}

\address[km]{Department of Physics and Astronomy,
Johns Hopkins University,
Baltimore, MD, USA}

    \begin{abstract}
      \noindent
It is known that the correlator of one axial and two vector currents, 
that receives leading contributions through one-loop fermion triangle 
diagrams,   is not modified by QCD radiative corrections at two loops. 
It was suggested that this non-renormalization 
of the $\langle VVA \rangle$ correlator persists in higher orders 
in perturbative QCD 
as well. To check this assertion, 
we compute the three-loop QCD corrections to the $\langle VAA \rangle $
correlator using the technique of asymptotic expansions.
We find that these corrections do not vanish and that they
are proportional to the  QCD $\beta$-function. 

      \end{abstract}

    \maketitle

\section{Introduction} 
The correlator  of two vector currents and one axial   current 
is an interesting  object.  Early studies 
of this correlator were important for developing 
an understanding 
that, in spite of naive equations of motion, 
 the axial current is anomalous and that the perturbative part of the anomaly 
can be computed {\em exactly} from the one-loop triangle diagram
\cite{Adler:1969er}.  The anomaly corresponds to just one 
term in the decomposition of the correlator into independent Lorentz
structures, and the non-renormalization of the anomaly 
by higher-order QCD corrections 
does not say if other terms share the same property. 
In fact, since the discovery of the non-renormalization 
of the anomaly, it was generally believed that 
coefficients of  other Lorentz structures change in higher orders 
of perturbation theory. This belief was challenged in 
Ref.~\cite{Vainshtein:2002nv},
where it was pointed out 
that  in the kinematic limit where momentum  of 
one of the vector currents is  vanishingly small, another 
non-renormalization theorem is valid. 
Indeed, in that limit just two independent form factors are 
needed to fully describe the
$\langle VVA \rangle $ 
correlator. One of these form factors 
is  the axial anomaly and, therefore, it 
is not renormalized. It was shown in   Ref.~\cite{Vainshtein:2002nv}  
that  due to  helicity conservation
in massless QCD, the two form factors are in fact proportional
to each other, and so the non-renormalization of 
one of them implies the non-renormalization of the other. 
Therefore, we come to the conclusion 
that the  {\it entire} $\langle VVA \rangle $  correlator 
is not renormalized in that limit. 

This result initiated  new studies of the $\langle VVA \rangle $ correlator, 
which led to  a number of surprising findings. First, 
in Ref.~\cite{Knecht:2003xy}  
it was pointed out that additional non-renormalization
theorems for form factors of the $\langle VVA \rangle $ correlator 
exist {\it even in the case of the most 
general kinematics}.  Later, in Ref.~\cite{Jegerlehner:2005fs} an
explicit calculation of the two-loop ${\cal O}(\alpha_s)$ 
contribution to the 
correlator was reported. The result turned out to be 
extremely simple: it was found that the 
${\cal O}(\alpha_s)$ correction to the correlator 
vanishes identically in   {\it the most general} kinematics
 if a consistent 
definition of the axial current is employed.  This unusual feature 
of radiative corrections  prompted the
authors of Ref.~\cite{Jegerlehner:2005fs} to speculate that their 
result might be an early indication of the non-renormalization of 
the entire $\langle VVA \rangle$ correlator 
in perturbative QCD with massless quarks 
to all orders in the strong coupling constant. 

The goal of this Letter  is to show that this speculation is not correct 
and that radiative corrections to the 
$\langle VVA \rangle$ correlator appear at the three-loop 
order in perturbative QCD. 
These three-loop  corrections are, however, peculiar in that they are 
explicitly proportional to the QCD $\beta$-function and 
{\it vanish} in the conformal limit of QCD. This is a natural 
result. In fact, it was pointed long ago in Ref.~\cite{schreier} 
that the {\it exact} 
$\langle VVA \rangle$ correlator  in a conformally-invariant theory 
is given by the one-loop expression and no contributions at higher-loops 
are allowed.  This happens because  conformal symmetry 
restricts the functional form of the three-point function up to 
a possible multiplicative factor which, however, is fixed 
by the requirement that the non-renormalization of the anomaly works 
out correctly.  It is easy to realize that through two loops in 
perturbative QCD, no diagrams that cause 
violation of the conformal symmetry 
contribute to  $\langle VV A \rangle $ correlator, while at three loops 
diagrams that correspond to the running of the coupling constant 
explicitly appear. This allows us to understand why 
no corrections to $\langle VVA \rangle $ 
correlator were found in Ref.~\cite{Jegerlehner:2005fs},  
why such corrections appear  at the three-loop order, 
and 
why they are proportional to the QCD $\beta$-function. 

The remainder of this Letter  is organized as follows.  In the next Section 
we define the $\langle VVA \rangle $ correlator and discuss 
its decomposition in terms of invariant form factors. In 
Section~\ref{threeloop} we explain  how three-loop 
contributions to the correlator can be computed. In Section~\ref{results} 
we present the results of the calculation. We conclude 
in Section~\ref{concl}.

\begin{figure}[t]
\begin{center}
\includegraphics[width=0.25\columnwidth]{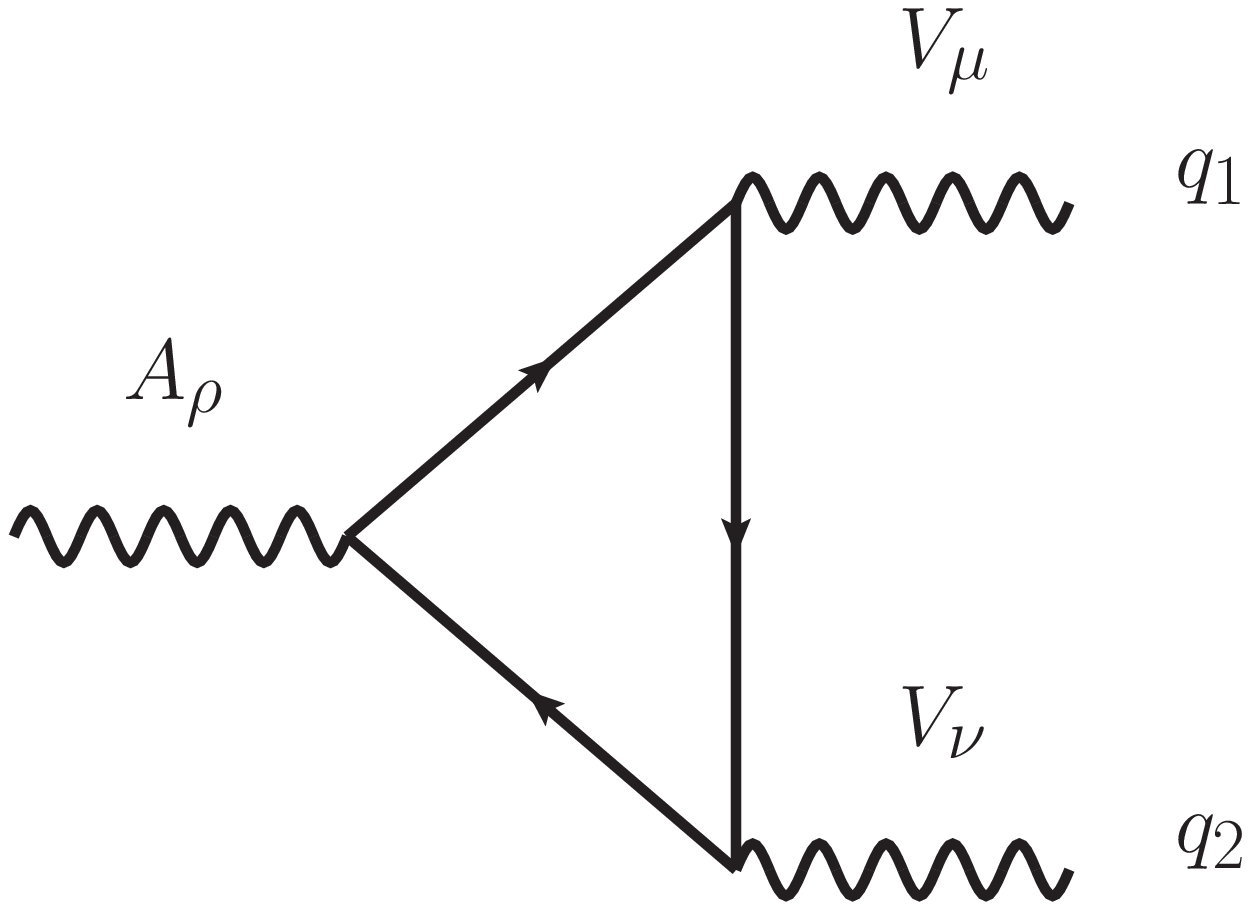}
\includegraphics[width=0.25\columnwidth]{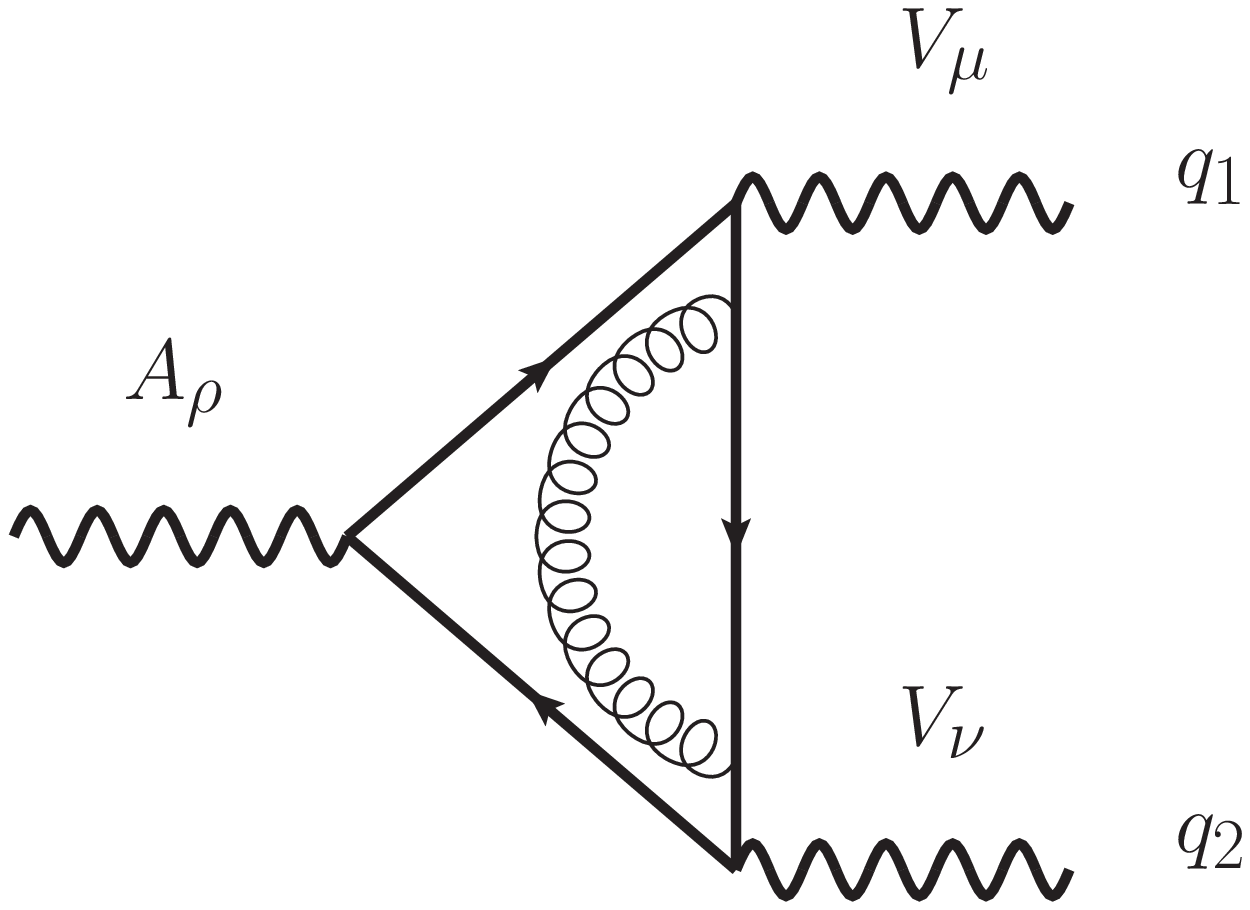}
\includegraphics[width=0.25\columnwidth]{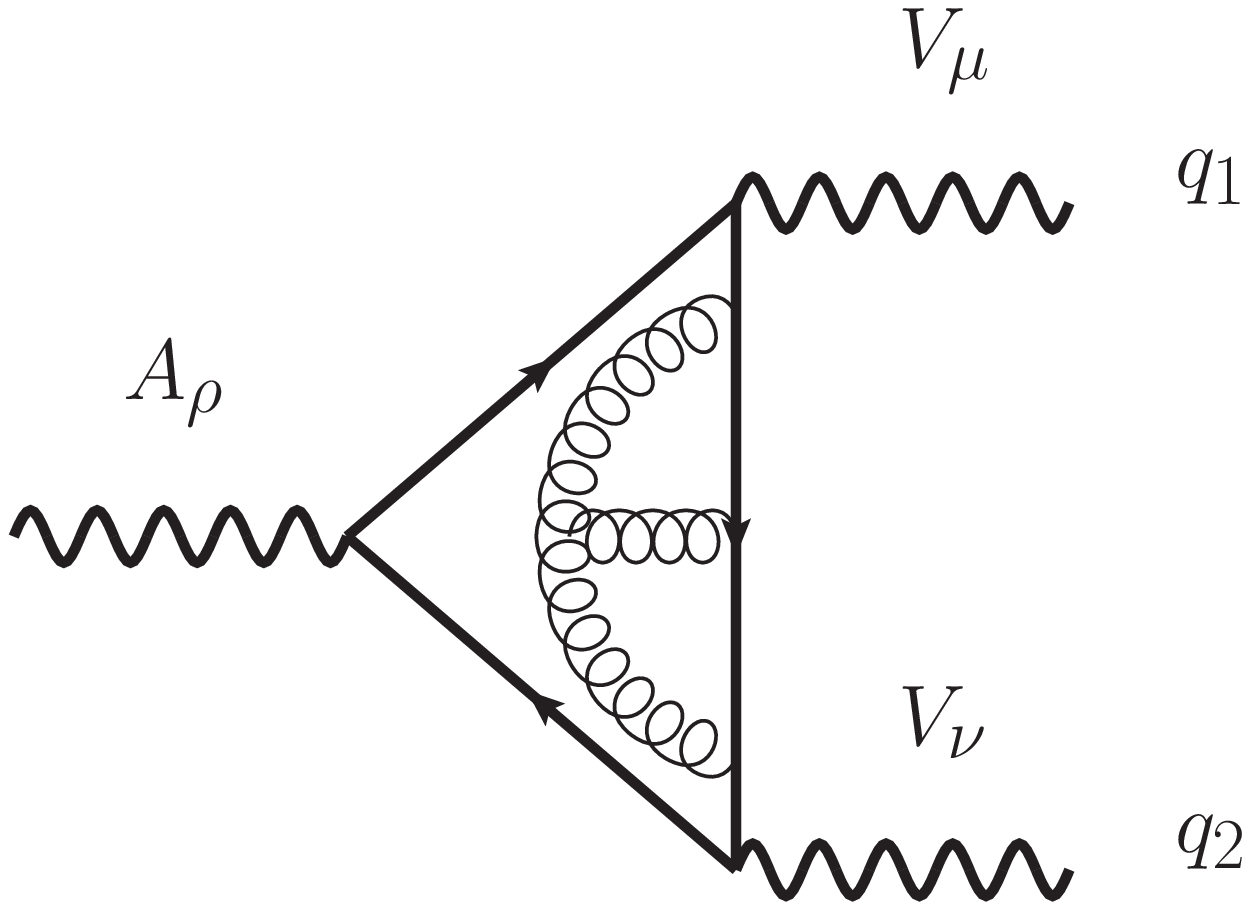}
\caption{Sample diagrams that contribute to $\langle VVA \rangle$ 
correlator at one, two and three-loops. The momenta $q_{1,2}$ are 
outgoing.
}
\label{fig1}
\end{center}
\end{figure}

\section{Definitions}
\label{defi}

We consider the correlator of two vector currents 
$V_\mu$ and one axial current $A_\mu$   
\be
\label{Greencalw}
{\cal W}_{\mu\nu\rho}(q_1,q_2) = -\int d^4x_1 d^4x_2
\,e^{i(q_1\cdot x_1 + q_2\cdot x_2)}
\,\langle\,0\,\vert\,\mbox{T}\{V_\mu(x_1)V_\nu(x_2)A_\rho(0)\}
\,\vert\,0\,\rangle.
\ee 
The currents are flavor-diagonal.\footnote{Note, however, 
that we do not include the so-called singlet contributions 
to the correlator when we compute it in perturbation theory.} They read 
\be V_{\mu} \,=\,{\overline\psi}\gamma_\mu\,\psi \quad ,
\quad A_{\mu} \,=\,
{\overline\psi}\gamma_\mu\gamma_5\,\psi\,,
\ee
where $\psi$ are fermion fields charged under 
the $SU(N_c)$ color group.

The $\langle VVA \rangle$ correlator in Eq.(\ref{Greencalw}) 
can be expressed in terms of independent Lorentz structures 
and invariant form factors \cite{Knecht:2003xy}.  
A suitable  parameterization reads 
\be
\label{eq_jgt}
\begin{split}
{\cal W}_{\mu\nu\rho}(q_1,q_2) & = w_L(q_1^2,q_2^2,q^2) t^{(1)}_{\mu \nu \rho} 
+ w_T^{(+)}(q_1^2,q_2^2,q^2) t^{(+)}_{\mu \nu \rho}
\\
& 
+ w_T^{(-)}(q_1^2,q_2^2,q^2) t^{(-)}_{\mu \nu \rho} 
+ \tilde{w}_T^{(-)}(q_1^2,q_2^2,q^2) {\tilde t}^{(-)}_{\mu \nu \rho}\,, 
\end{split} 
\ee
where, following Ref.~\cite{Knecht:2003xy},
 we introduced four independent  tensor structures 
\be
\begin{split} 
& t^{(1)}_{\mu \nu \rho} = 
- q_{\rho}\epsilon_{\mu\nu\alpha\beta}q_1^{\alpha}q_2^{\beta}\,, \\
& t^{(+)}_{\mu \nu \rho} = 
q_{1\nu}\epsilon_{\mu\rho\alpha\beta}q_1^{\alpha}q_2^{\beta} 
- q_{2\mu}\epsilon_{\nu\rho\alpha\beta}q_1^{\alpha}q_2^{\beta}
- q_1\cdot q_2\epsilon_{\mu\nu\rho\alpha}\kappa^{\alpha}
-2 \frac{q_1 \cdot q_2 }{q^2}\epsilon_{\mu\nu\alpha\beta}q_1^{\alpha}q_2^{\beta} q_{\rho}\,,
\\ 
& t^{(-)}_{\mu \nu \rho} = 
\epsilon_{\mu\nu\alpha\beta}q_1^{\alpha}q_2^{\beta}\kappa_\rho
- \frac{\kappa \cdot q}{q^2}
\epsilon_{\mu\nu\alpha\beta}q_1^{\alpha}q_2^{\beta} q_{\rho}\,,
\\
& {\tilde t}^{(-)}_{\mu \nu \rho} = 
q_{1\nu}\epsilon_{\mu\rho\alpha\beta}q_1^{\alpha}q_2^{\beta} 
+ q_{2\mu}\epsilon_{\nu\rho\alpha\beta}q_1^{\alpha}q_2^{\beta}
 - q_1\cdot q_2
\epsilon_{\mu\nu\rho\alpha}q^{\alpha}\,,
\end{split} 
\ee
and momenta $q = q_1 + q_2,\; \kappa = q_1 - q_2$. From momentum 
conservation, $q$ is 
the incoming  momentum carried by the axial current. We note that 
all tensor structures 
are transversal with respect to the momenta of the 
vector currents
\be
q_1^{\mu} t_{\mu \nu \rho} = 0,\;\;\; q_2^\nu  t_{\mu \nu \rho} = 0\,, 
\ee
where $t_{\mu \nu \rho}$ 
is a generic notation for $t^{(1)}_{\mu \nu \rho},t^{(\pm)}_{\mu \nu \rho}$ 
and  ${\tilde t}^{(+)}_{\mu \nu \rho}$. On the other hand, 
these tensor structures have different transversality properties 
with respect to  the momentum $q$ of the axial current, 
\be
q^\rho t^{(1)}_{\mu \nu \rho} = -q^2 \epsilon_{\mu \nu \alpha \beta} q_1^{\alpha} q_2^{\beta}\,,
\;\;\;\; q^{\rho} t^{(\pm)}_{\mu \nu \rho} = 0\,,
\;\;\;\; q^{\rho} {\tilde t}^{(-)}_{\mu \nu \rho} = 0\,.
\ee
Therefore, the parameterization in Eq.(\ref{eq_jgt}) is consistent with the conservation 
of the vector current and with the fact that the conservation of the axial current is violated 
by triangle loop diagrams.  We will refer to $w_L$ as the 
longitudinal form factor 
and to $w_T^{(\pm)}$ and ${\tilde w}_T^{(-)}$ as 
the transversal form factors. 

We now discuss what is known about these  form factors. In general, 
they  are functions of three independent 
kinematic variables, $q_1^2, q_2^2$ and  $q^2 = (q_1 + q_2)^2$.  
Since the divergence of the axial current is fixed by the anomaly equation 
\be
\partial_\mu A^\mu=
\frac{N_c}{16\pi^2}\epsilon_{\mu \nu \rho \sigma}F^{\mu \nu}F^{\rho\sigma}\,, 
\label{eq_anom}
\ee
to all orders in perturbation theory, the longitudinal form factor 
$w_L$ is completely defined. Indeed, the anomaly equation 
implies 
\be
q^{\rho}{\cal W}_{\mu\nu\rho} = -w_L(q_1^2,q_2^2,q^2) 
q^2 \epsilon_{\mu \nu \alpha \beta} q_1^{\alpha} q_2^{\beta} 
= -\frac{i N_c }{2\pi^2} \epsilon_{\alpha\beta\mu\nu}q_1^{\alpha}q_2^{\beta}\,.
\ee
We solve this equation for $w_L$ to find 
\be
\label{longeq}
 w_L(q_1^2,q_2^2,q^2) = \frac{i\,N_c}{2\pi^2 q^2}\,.
\ee

While the longitudinal form factor can be fully determined from the 
anomaly equation, 
the transversal form factors can not.  However, as it was shown in 
Refs.~\cite{Vainshtein:2002nv,Knecht:2003xy}, 
some combinations of transversal  form factors are uniquely fixed 
by the chiral symmetry in perturbative QCD with massless flavors.   
The following relations are valid~\cite{Knecht:2003xy},
\be
\label{exact_res}
\begin{split} 
& \left [ w_T^{(+)} + w_T^{(-)} \right ](q_1^2,q_2^2,q^2)
-\left [ w_T^{(+)} + w_T^{(-)} \right ](q^2,q_2^2,q_1^2) = 0\,,
\\
& \left [ {\tilde w}_T^{(-)} + w_T^{(-)} \right ](q_1^2,q_2^2,q^2)
+\left [ {\tilde w}_T^{(-)} + w_T^{(-)} \right ](q^2,q_2^2,q_1^2) = 0\,,
\\
& \left [ w_T^{(+)} + {\tilde w}_T^{(-)} \right ](q_1^2,q_2^2,q^2)
+
\left [ w_T^{(+)} + {\tilde w}_T^{(-)} \right ](q^2,q_2^2,q_1^2)
-
w_L(q^2,q_2^2,q_1^2)
\\
&=
- \left[ 2 \frac{q_2 \cdot q}{q_1^2} w_T^{(+)}(q^2,q_2^2,q_1^2)
- 2\frac{q_1 \cdot q_2}{q_1^2}w_T^{(-)}(q^2,q_2^2,q_1^2)\right]\,.
\end{split} 
\ee

In perturbation theory, the correlator  ${\cal W}_{\mu \nu \rho }$ 
receives contributions starting at one-loop order, see 
Fig.~\ref{fig1}.   The correlator was computed through 
two loops in Ref.~\cite{Jegerlehner:2005fs} for 
arbitrary momenta $q_1$ and $q_2$.
   The results of that calculation 
showed that, with the proper definition 
of the axial current,  the two-loop corrections to the {\it full} 
correlator vanish. We pointed out a possible reason for that 
in Section~1, where we noted 
that it is natural to expect that these results will be violated 
starting from three loops. In the next Section we discuss 
how to set up a calculation to check this explicitly.

\section{The three-loop calculation}
\label{threeloop}

We are interested in computing the ${\cal W}_{\mu \nu \rho}$
 correlator to three loops  in perturbative QCD, for {\it arbitrary} 
$q_1$ and $q_2$.   Sample  diagrams are shown in Fig.~\ref{fig1}. 
We note that the one-loop contribution is described by 2 diagrams, 
the two-loop contribution by 12  diagrams and 
the three-loop contribution by 182 diagrams.
It goes without saying that, at present, no computational technology 
exists that can be used to calculate three-loop 
three-point functions with three 
external off-shell legs.  To circumvent this problem, we consider 
a hierarchy of incoming momenta,  $q_2 \ll q_1$, and construct a systematic 
expansion of the three-loop diagrams in the ratio $q_2/q_1$.  It is 
intuitively clear that, upon such an expansion, a three-point function 
is mapped onto a set of two-point functions and its derivatives 
evaluated either for  small $k \sim q_2$ or large $k \sim q_1$ 
loop momenta. 
Such an expansion is 
constructed  most efficiently by 
using standard techniques of the large momentum expansion 
\cite{smirnov}. In the current application, we use 
this procedure as automated in the computer programs  \texttt{q2e} and
\texttt{exp}~\cite{Harlander:1997zb,Seidensticker:1999bb}.
Once the three-loop three-point functions are mapped onto the  
three-loop  two-point functions by the large momentum expansion 
procedure, we use the program 
\texttt{MINCER}~\cite{MINCER} to compute them. 
Finally, we note that all diagrams that contribute
to the correlator are generated with 
\texttt{Qgraf}~\cite{Nogueira:1991ex}.

As a technical remark, we note that 
treatment of tensor integrals that appear 
in this computation requires some care.  
At one  loop, \texttt{MINCER} routines can deal with tensor integrals
of arbitrary rank, contracted   with arbitrary  external momenta. 
Unfortunately,  at higher loops \texttt{MINCER} can only be used 
to compute  two-point scalar integrals  that depend 
on a single external momentum.  However, since 
we arrive at relevant 
two-point functions by  expanding three-point functions in
$q_2/q_1$, we need to compute  more complicated tensor integrals.
To give an example,  consider an integral 
\be
\int \prod \limits_{i=1}^{3} 
\frac{{\rm d}^d k_{i}}{(2\pi)^d} 
\frac{ (k_1 \cdot q_2)^{a_1} (k_2 \cdot q_2)^{a_2}..  
}{k_1^2 k_2^2 ...  (k_1 - q_1)^2..(k_2-q_1)^2..},
\label{eq_form}
\ee
where  the  two-point function itself 
depends on momentum $q_1$ but   the numerator contains scalar products 
of loop momenta $k_i$ with another external momentum $q_2$. This is 
a typical situation for us to face, 
and it can not be handled by \texttt{MINCER}.
To deal  with the integrals of the type shown in 
Eq.(\ref{eq_form}), we write 
$
q_2^{\mu} = (q_2 \cdot q_1)/q_1^2 q_1^\mu + q_{2,\perp}^{\mu},
$
where $q_{2,\perp} \cdot q_1 = 0$,  and note  that 
the final result can not depend on the direction of $q_{2,\perp}$.  
We make use of this observation 
by {\it averaging } over directions of $q_{2,\perp}$ 
in Eq.(\ref{eq_form}). Such averages are constructed 
from  the metric tensor $g^{\mu \nu} -q_1^\mu q_1^\mu/q_1^2$ 
which only contains vector $q_1$ and, therefore, leads 
to scalar products of four-momenta 
 that can be treated by \texttt{MINCER}.

Computational  
procedures described above rely heavily on the use 
of dimensional regularization. This leads to a subtlety 
since, as it is well known, 
continuation of the Dirac matrix $\gamma_5$ to $d$ dimensions
requires care and a proper way to do so is crucial 
for the correct 
computation of the $\langle VVA \rangle $ correlator. 
 We use the definition of the axial current 
developed in Ref.~\cite{Larin:1993tq} which, on one hand, is self-consistent 
and, on the other hand, does not introduce unnecessary difficulties 
into already complicated multi-loop computation. Thus, we define the axial 
current in the following way~\cite{Larin:1993tq},
\be
\label{defA}
A_{\rho} \equiv 
\frac{i}{6}\epsilon_{\rho\alpha' \beta' \lambda '}
{\bar \psi} \Gamma^{[\alpha' \beta' \lambda' ]} 
\psi, \quad \text{with}\,\,\,\epsilon_{0123}=1=-\epsilon^{0123}\,,
\ee
where 
$\Gamma^{\alpha \beta \lambda}= \gamma^\alpha \gamma^\beta \gamma^\lambda$
and 
square brackets indicate the anti-symmetrization of the Lorentz 
indices,
\be
\label{brackets}
\Gamma^{[\alpha \beta \lambda']}=
\frac{1}{6} \left ( 
 \gamma^{\alpha } \gamma^{\beta } \gamma^{\lambda }
-\gamma^{\alpha } \gamma^{\lambda} \gamma^{\beta}+\dots 
\right )\,.
\ee
With this definition, the axial current needs renormalization. We write 
$A_{\rho} = Z_{A} A_{ \rho}^{\rm bare}$ and note that 
the renormalization constant $Z_A$ was worked out in 
 Ref.~\cite{Larin:1993tq}. There, 
in addition to the $\overline{\rm MS}$ renormalization constant, 
a finite piece is added that ensures that the 
non-singlet current has no anomalous dimension and the 
anomaly equation (\ref{eq_anom}) is 
satisfied. The full renormalization constant reads at two loops
\begin{equation}
\begin{split} 
Z_{A}= & \left (  1 + \frac{2 a_s^2C_F}{\epsilon}\beta_0
 \right ) \times \left (  1 - 4C_Fa_s 
+ C_F a_s^2\left ( 22\,C_F - \frac{107}{9}\,C_A + \frac{2}{9}\,n_f \right )
 \right )\,,
\label{ren_a}
\end{split} 
\end{equation}
where $a_s=g^2(\mu)/(16\pi^2)$ is the $\overline {\rm MS}$ QCD coupling 
constant defined at the scale $\mu$ and 
$\beta_0 = 11/3C_A - 2/3n_f$ is the one-loop QCD $\beta$-function.
Also,   $n_f$ is the number of (massless) quark flavors 
and $C_F=(N_c^2-1)/(2N_c)$, $C_A=N_c$ 
are the  Casimir operators of the ${\rm SU}(N)$ 
group in fundamental and adjoint representations, respectively. 
In QCD, we use $N=3$.

We now return to the definition of the axial current in Eq.~(\ref{defA})  and 
observe that it contains the Levi-Civita tensor $\epsilon_{\mu \nu \alpha \rho}$, which is a 
four-dimensional object whose continuation to $d$ dimensions is not possible. 
To address this issue, we first compute an auxiliary tensor correlator 
${\cal T}_{\alpha\beta\lambda,\mu\nu}$  which does  not contain  the Levi-Civita tensor
and which is finite at $d=4$. We then obtain the required correlator by 
contracting ${\cal T}_{\alpha\beta\lambda,\mu\nu}$ with the Levi-Civita tensor
\be
\label{eqwmunu}
{\cal W}_{\mu\nu\rho}=\frac{i}{6}\epsilon_{\rho\alpha\beta\lambda}{\cal T}_{\alpha\beta\lambda,\mu\nu}\,.
\ee
This  can be easily done because 
all entries on the right hand side of Eq.(\ref{eqwmunu}) are well-defined 
in four dimensions.

To define the auxiliary tensor ${\cal T}_{\alpha\beta\lambda,\mu\nu}$, we 
imagine 
that the correlator ${\cal W}_{\mu\nu\rho}$ is contracted 
with $\epsilon_{\alpha \beta \lambda \rho}$ 
\be
{\cal T}_{\alpha\beta\lambda,\mu\nu}(q_1,q_2) \equiv  \frac{1}{i}\epsilon_{\alpha\beta\lambda\rho}{\cal W}_{\mu\nu\rho}(q_1,q_2)\,.
\ee
We then use the identity 
\be
\epsilon_{\alpha\beta\lambda\rho}\epsilon_{\rho\alpha'\beta'\lambda'} = 
g_{\alpha,\alpha'} g_{\beta,\beta'} g_{\lambda,\lambda'} 
- g_{\alpha,\alpha'} g_{\beta,\lambda'} g_{\lambda,\beta'} + \dots
\ee
to remove  the Levi-Civita tensors in favor of  
products of metric tensors and find a convenient 
expression for the auxiliary correlator 
\be
\begin{split} 
\label{Tdef}
& {\cal T}_{\alpha \beta \lambda,\mu\nu}(q_1,q_2) = 
-\frac{1}{6} 
\left ( 
g_{\alpha \alpha'} g_{\beta \beta'} g_{\lambda \lambda'} 
- g_{\alpha \alpha'} g_{\beta \lambda'} g_{\lambda \beta'} 
+..
\right )\\
& \times \int d^4x_1 d^4x_2
\,e^{i(q_1\cdot x_1 + q_2\cdot x_2)}
\,\langle\,0\,\vert\,\mbox{T}\{V_\mu(x_1)V_\nu(x_2) \bar \psi(0) 
\Gamma^{[\alpha' \beta' \lambda']} \psi(0)\}
\,\vert\,0\,\rangle\,.
\end{split} 
\ee 
Because the right-hand side of Eq.(\ref{Tdef}) can be analytically continued to 
$d$ dimensions in a straightforward way, 
the correlator ${\cal T}_{\alpha\beta\lambda,\mu\nu}(q_1,q_2)$ 
can be calculated in dimensional regularization. After the  renormalization 
of the ``axial'' current Eq.(\ref{ren_a}) is applied, 
finite result 
for ${\cal T}_{\alpha\beta\lambda,\mu\nu}(q_1,q_2)$ is obtained. 
Finally,  we use Eq.(\ref{eqwmunu}) to calculate 
the correlator ${\cal W}_{\mu \nu \rho}$. 

Before discussing the results of the computation, it is useful 
to comment on  the  Lorentz decomposition of the correlator 
${\cal T}_{\alpha\beta\lambda,\mu\nu}$. As follows from its definition, Eq.(\ref{Tdef}), 
${\cal T}_{\alpha\beta\lambda,\mu\nu}(q_1,q_2)$ is an anti-symmetric tensor with respect to 
its first three indices and it 
is a symmetric tensor under a simultaneous change 
$q_1 \leftrightarrow q_2$, $\mu \leftrightarrow \nu$. We find that 
it is possible to express ${\cal T}$ in terms of six tensor structures
\be
{\cal T}_{\alpha\beta\lambda,\mu\nu}(q_1,q_2) 
= \sum_{i=1}^6 c_i(q_1,q_2)\,T^{(i)}_{\alpha\beta\lambda,\mu\nu}\,,
\label{tens}
\ee
where 
\be
\begin{split} 
\label{defT}
& T^{(1)}_{\alpha\beta\lambda,\mu\nu}=[q_{1\alpha} g_{\beta\mu}g_{\lambda\nu}]\,,
\;\;\;\;\;\;\;
T^{(2)}_{\alpha\beta\lambda,\mu\nu}=[q_{2\alpha} g_{\beta\mu}g_{\lambda\nu}]\,,
\;\;\;\;\;\;\;
T^{(3)}_{\alpha\beta\lambda,\mu\nu}=[q_{1\alpha}q_{2\beta}g_{\mu\lambda}q_{1\nu}]\,,
\\
&  T^{(4)}_{\alpha\beta\lambda,\mu\nu}=[q_{1\alpha}q_{2\beta}g_{\nu\lambda}q_{1\mu}]\,,
\;\;\;
T^{(5)}_{\alpha\beta\lambda,\mu\nu}=[q_{1\alpha}q_{2\beta}g_{\mu\lambda}q_{2\nu}]\,,
\;\;\;
T^{(6)}_{\alpha\beta\lambda,\mu\nu}=[q_{1\alpha}q_{2\beta}g_{\nu\lambda}q_{2\mu}]\,.
\end{split} 
\ee
The square brackets here imply an 
anti-symmetrization with respect to indices $\alpha$, $\beta$ and $\lambda$ while the
positions of $\mu$ and $\nu$ are kept fixed. 

While the tensor ${\cal T}_{\alpha \beta \lambda, \mu \nu}$ can be conveniently computed using the asymptotic 
expansion technique, some care may be required with the interpretation 
of the result, since the coefficients $c_{1-6}(q_1,q_2)$ 
are not independent.  Relations 
between them can be obtained 
from the conservation of the vector  currents and from 
the non-renormalization of the 
axial anomaly.  We find three relations 
\begin{equation} 
\begin{split} 
 c_2 - q_1^2\,c_4 - q_1\cdot q_2\,c_6 =0\,,\;\;\;
 c_1 - q_1\cdot q_2\,c_3 - q_2^2\,c_5 =0\,,\;\;\;
c_1 - c_2= \frac{N_c}{2\pi^2}\,. 
\end{split} 
\label{eqsc}
\ee
In the actual three-loop computation, Eqs.(\ref{eqsc}) are 
not enforced but are used as checks of its correctness. Once coefficients 
$c_{1-6}$ are known, we compute    ${\cal W}_{\alpha \beta \lambda}$ 
using Eq.(\ref{eqwmunu}) and re-write it through the form factors 
$w_T^{(\pm)}, {\tilde w}_T^{(-)}$ and $w_L$. The corresponding results 
are presented in the next Section.

\section{Results} 
\label{results}

We are now in
 position to present the results of the calculation. To this 
end, 
we introduce  the notation 
\be
r_1 = \frac{q_1 \cdot q_2}{q_1^2},\;\;\;r_2 = \frac{q_2^2}{q_1^2},\;\;\; L_r = \ln r_2\,.
\ee
We perform the calculation in the limit $q_2/q_1 \ll 1$, 
so that $r_1^2 \sim r_2 \ll 1$. The results 
for the form factors below are given as an expansion in $r_{1,2}$.
For each  transversal form factor we find that 
\begin{itemize} 
\item the two-loop QCD corrections 
vanish, in accord with the observation of 
Ref.~\cite{Jegerlehner:2005fs};
\item the three-loop QCD corrections {\it do not } vanish, but are 
proportional to the  one-loop QCD $\beta$-function.\footnote{We note 
that the simplest way to check that the three-loop corrections 
do not vanish is to study $n_f$-dependent vacuum polarization 
insertion   diagrams. 
Since the correction to the correlator turns out to be proportional
to the $\beta$-function, such diagrams give, in fact, the full answer.}
  As a result, 
they vanish in the conformal $\beta \to 0$ limit.
\end{itemize}

The longitudinal form factor is known exactly from the anomaly 
equation; it is given in Eq.~(\ref{longeq}). We  do not discuss it anymore. 
The perturbative expansion for a generic transversal form factor 
$w_T = w_{T}^{\pm}, {\tilde w}_T^{-}$ is written as 
\be
w_{T}(q_1^2,q_2^2,q^2) = \frac{iN_c}{16\pi^2 q_1^2} 
\left (  
w_{T}^{(1)} + \beta_0 C_F 
\left ( \frac{\alpha_s(\mu)}{4 \pi} \right )^2 w_{T}^{(3)} 
+{\cal O}(\alpha_s^3) \right )\,,
\ee
where $\beta_0 = 11/3 C_A - 2/3 n_f$.
We calculated $w_T^{(1,3)}$ using asymptotic expansions in $q_2/q_1$ through 
fourth order.  Upon expanding the one-loop expressions for these 
form factors presented in Ref.~\cite{Jegerlehner:2005fs}, we find 
full agreement with our result.\footnote{We note that this agreement 
is only found if  we multiply all
results   in  Ref.~\cite{Jegerlehner:2005fs} by a factor two that, we believe, 
is  erroneously omitted there.} 

 Below we present  the expansions of the  one-loop and three-loop 
form factors through second order in $q_2/q_1$. 
For the one-loop contributions we find 
\be
\label{woneloop}
\begin{split} 
& w_T^{(+,1)} = \frac{20}{9} -  \frac{4}{3}L_r 
-\frac{2r_1}{9} \left ( 1-6 L_r \right ) - 
r_1^2 \left (\frac{52}{75} + \frac{8}{5}L_r \right) + r_2 \left(\frac{82}{225} + \frac{8}{15} L_r \right )
+{\cal O}(r_1^3)\,, 
\\
& w_T^{(-,1)} = -\frac{16}{9} -  \frac{4}{3}L_r 
+\frac{2r_1}{9} \left ( 11+6 L_r \right ) - 
r_1^2 \left ( \frac{84}{25} + \frac{8}{5}L_r \right) 
+ r_2 \left(\frac{48}{25} - \frac{4}{5} L_r \right )
+{\cal O}(r_1^3)\,,
\\
& {\tilde w}_T^{(-,1)} = -w_T^{(-,1)} + {\cal O}(r_1^3)\,.
\end{split}
\ee

The three-loop contributions  read
\be
\label{wthreeloops}
\begin{split} 
 w_T^{(+,3)}
& = 2 L_r^2 + (-14+16 \zeta_3) L_r + \frac{94}{3} - \frac{56}{3}\zeta_3
-r_1 \left ( 2 L_r^2 - (10-16 \zeta_3) L_r + \left ( 16 - \frac{8 \zeta_3}{3} \right ) \right ) 
\\
& +r_1^2 \left ( \frac{3853}{270} +\frac{128}{25} \zeta_3 
 - \left ( \frac{91}{9} 
- \frac{96}{5} \zeta_3 \right ) L_r  + \frac{22}{9} L_r^2 \right ) 
\\
& + r_2 \left ( - \frac{6991}{540} - \frac{88}{75}\zeta_3 
+ \left ( \frac{125}{18} - \frac{32}{5} \zeta_3 \right ) L_r - \frac{11}{9}L_r^2 
\right ) + {\cal O}(r_1^3)\,, \\
 w_T^{(-,3)}
 & = 2 L_r^2 + (-14+16 \zeta_3) L_r + \frac{94}{3} - \frac{56}{3}\zeta_3
-r_1\left(2L_r^2 + (-10+16\zeta_3)L_r +\frac{56}{3}-\frac{8}{3} \zeta_3\right)
\\
& +r_1^2 \left ( \frac{4573}{270} +\frac{128}{25} \zeta_3 
 - \left ( \frac{91}{9} 
- \frac{96}{5} \zeta_3 \right ) L_r  + \frac{22}{9} L_r^2 \right ) 
\\
& + r_2 \left ( \frac{9209}{540} - \frac{496}{25}\zeta_3 
- \left ( \frac{127}{18} - \frac{48}{5} \zeta_3 \right ) L_r + \frac{7}{9}L_r^2 
\right ) + {\cal O}(r_1^3)\,,
\\
{\tilde w}_T^{(-,3)} &  = - w_T^{(-3)}+ {\cal O}(r_1^3)\,.
\end{split} 
\ee

It is now straightforward  to check the exact 
perturbative QCD relations between different form factors, 
shown in Eq.(\ref{exact_res}). 
The above expansions are given for $w_T(q_1^2,q_2^2,q^2)$; 
to compute Eq.(\ref{exact_res}), we require $w_T(q^2,q_2^2,q_1^2)$.
These can be easily obtained from the above equations by applying 
the following transformations to them 
\be
q_1^2 \to q_1^2(1 + 2 r_1 + r_2),\;\;\;
r_2 \to r_2 /(1+2 r_1 + r_2),
\;\;\; r_1 \to -\frac{r_1 + r_2}{1+2r_1+r_2}\,.
\ee
Upon applying these transformations to 
Eqs.(\ref{woneloop},\ref{wthreeloops}),  
re-expanding in $r_1^2 \sim r_2 \ll 1$  and combining the form factors 
in the right way,  we 
find that Eqs.(\ref{exact_res}) 
are indeed satisfied.   As a simple illustration of this fact, 
we note \cite{Knecht:2003xy}  
that, in the limit $q_2 \to 0$, these equations reduce to
\be
w_L(q_1^2,0,q_1^2) = 2 \left ( w_T^{(+)}(q_1^2,0,q_1^2) 
 + {\tilde w}_T^{(-)}(q_1^2,0,q_1^2) \right )\,.
\ee
Since the 
longitudinal form factor does not receive QCD corrections, the three-loop 
corrections to $w_T^{(+)}$ and $w_T^{(-)}$ in these kinematics must cancel. 
Taking  the limit $r_2 \to 0, r_1 \to 0$ in Eq.(\ref{wthreeloops})
we observe the required cancellation.

\section{Conclusions} 
\label{concl}

In this Letter, we described  the calculation of the three-loop 
perturbative QCD contribution to the 
correlator of 
one axial and two vector currents for general kinematics. In perturbative QCD 
this correlator receives contributions starting at the one-loop 
order. The two-loop contribution 
was computed in  Ref.~\cite{Jegerlehner:2005fs}, where  it was observed 
that, for a propely defined axial current, the two-loop contribution vanishes. 
This result prompted the authors of Ref.~\cite{Jegerlehner:2005fs} 
to suggest that this non-renormalization of the {\it full} correlator will 
carry over to yet higher orders  in perturbative QCD.

Our three-loop computation clarifies this issue. We compute the three-loop 
contributions to the correlator in arbitrary kinematics, using the techniques 
of asymptotic expansions. We find that the 
three-loop contributions do not vanish
but that they are proportional to the QCD $\beta$-function.  This feature 
explains the vanishing of the two-loop contribution to $\langle VVA \rangle $ 
as being due to  conformal symmetry of QCD. Indeed, it was pointed out
in Ref.~\cite{schreier} that in the conformally-invariant theory 
the functional form of the correlator is fixed. 
Through two-loops the perturbative contributions 
to  $\langle VVA \rangle $ in perturbative QCD are not sensitive 
to the breaking of conformal symmetry, but  this changes at three 
loops because diagrams that describe the  running of the coupling 
constant appear for the first time. 
Thus, the three-loop  contribution should be proportional 
to the $\beta$-function, in full accord with our 
explicit computation.  Finally, we have verified the validity 
of all non-renormalization  theorems in perturbative 
QCD for the form factors of the $\langle VVA \rangle $ correlator
derived in   Refs.~\cite{Vainshtein:2002nv,Knecht:2003xy}. We find 
that through three-loops, all these theorems are satisfied.

{\bf Acknowledgments} 
K.M. is grateful to A.~Vainshtein for many enlightining 
conversations  about Ref.~\cite{Jegerlehner:2005fs}.
We are indebted to K.G.~Chetyrkin for discussions and 
his help with the calculation described in this paper. 
The  research of K.M. is partially supported by US NSF under grants PHY-1214000
and by Karlsruhe Institute of Technology through 
is distinguished researcher fellowship program.
The research of J.M. is supported by  the Deutsche Forschungsgemeinschaft in the
Sonderforschungsbereich/Transregio SFB/TR9 ``Computational Particle Physics".


\begin{thebibliography}{99}


\bibitem{Adler:1969er}
  S.~L.~Adler and W.~A.~Bardeen,
  Phys.\ Rev.\  {\bf 182} (1969) 1517.
  

\bibitem{Vainshtein:2002nv} 
  A.~Vainshtein,
  Phys.\ Lett.\ B {\bf 569}, 187 (2003)
  [hep-ph/0212231].



\bibitem{Knecht:2003xy} 
  M.~Knecht, S.~Peris, M.~Perrottet and E.~de Rafael,
  JHEP {\bf 0403}, 035 (2004)
  [hep-ph/0311100].


\bibitem{Jegerlehner:2005fs}
  F.~Jegerlehner and O.~V.~Tarasov,
  Phys.\ Lett.\ B {\bf 639} (2006) 299
  [hep-ph/0510308].

\bibitem{schreier} 
E.J.~Schreier, Phys. Rev.~{\bf D} 3  (1971), 980.


\bibitem{Larin:1993tq}
  S.~A.~Larin,
  Phys.\ Lett.\ B {\bf 303} (1993) 113
  [hep-ph/9302240].

\bibitem{Nogueira:1991ex}
  P.~Nogueira,
  J.\ Comput.\ Phys.\ \textbf{105}, 279 (1993).

\bibitem{smirnov} 
V.A.~Smirnov, {\it Renormalization and Asymptotic expansions}
(Birkh\"ausen, Basel, 1991).

\bibitem{Harlander:1997zb}
  R.~Harlander, T.~Seidensticker and M.~Steinhauser,
  Phys.\ Lett.\ B \textbf{426}, 125 (1998) [arXiv:hep-ph/9712228].


\bibitem{Seidensticker:1999bb}
  T.~Seidensticker,
  arXiv:hep-ph/9905298.

\bibitem{Vermaseren:2000nd}
  J.~A.~M.~Vermaseren,
  arXiv:math-ph/0010025.

\bibitem{MINCER}
S.~A.~Larin, F.~V.~Tkachov, J.~A.~M.~Vermaseren,
Rep.~No.~NIKHEF-H/91-18, Amsterdam, 1991.


\end{thebibliography}
\end{document}